\documentstyle[12pt]{article}

\begin{document}

\begin{titlepage}

\begin{flushright}
{\sc TRI-PP-94-78}\\ [.2in]
{\sc September, 1994}\\ [.5in]
\end{flushright}

\begin{center}
{\LARGE
``Non-factorizable'' terms in hadronic B-meson weak decays.}\\ [.5in]
{\large
Jo\~{a}o M. Soares}\\ [.1in]
{\small
TRIUMF, 4004 Wesbrook Mall, Vancouver, BC Canada V6T 2A3}\\ [.5in]

{\normalsize\bf
Abstract}\\ [.2in]
\end{center}

{\small
The branching ratios for the hadronic B-meson weak decays
$B \rightarrow J/\psi K$ and $B \rightarrow D \pi$ are used to extract the size
of the ``non-factorizable'' terms in the decay amplitudes. It is pointed out
that the solutions are not uniquely determined. In the $B \rightarrow J/\psi K$
case, a 2-fold ambiguity can be removed by analyzing the contribution of this
decay to $B \rightarrow K l^+ l^-$. In the $B \rightarrow D \pi$ case, a 4-fold
ambiguity can only be removed if the ``non-factorizable'' terms are assumed
to be a small correction to the vacuum insertion result.
}

\end{titlepage}

\section{Introduction}

An increasing sample of $B$-mesons has been gathered from different
experiments in recent times, and will tend to increase sharply in the
near future with the advent of $B$-factories and, possibly, experiments
in hadron colliders targeted at $B$-physics. The major concern in this
paper is to clarify how such wealth of data can be used to study some
of the aspects that remain unclear in the hadronic weak decays of the
$B$ and the other flavored mesons. The focus shall be on the 2-body decays
that proceed through the tree level, Cabibbo favored, quark transitions
$b \rightarrow c\overline{c}s$ and $b \rightarrow c\overline{u}d$. The
corresponding effective weak Hamiltonian, once QCD corrections have been
included, is the sum of two 4-quark operators, that only differ in the color
indices of their quark fields and the strength of their Wilson coefficients.
In the calculation of the decay amplitudes, one is faced with the task of
evaluating the matrix elements, between the initial and final hadronic states,
of those two 4-quark operators. The vacuum insertion (factorization)
approximation \cite{Cheng:89} reduces this problem to that of determining
the matrix elements of bilinear quark operators; such matrix elements can
then be measured from leptonic and semi-leptonic decays, or calculated in
some model for the mesons. Unfortunately, this is a poor approximation for
one of the 4-quark operators: due to a mismatch in the color indices, its
matrix element has significant ``non-factorizable'' terms \cite{Cheng:89},
that do not appear in the vacuum insertion approximation. That these
terms are important can be seen, for example, in the strong disagreement
between the factorization predictions and the observed rates for the color
suppressed $B$ or $D$ decays (see, for example, ref. \cite{BSW:87}. In the
color suppressed decays, the effect of the ``non-factorizable'' terms is
enhanced by an accidental cancellation of the terms that are non-zero in the
vacuum insertion approximation).

The standard procedure \cite{BSW:87} in dealing with this discrepancy has been
to preserve the vacuum insertion result for the hadronic matrix elements, but
to replace the Wilson coefficients that multiply them by two free parameters.
These parameters are then determined from a fit to the observed values of the
branching ratios. For the case of the $D$- and $B$-meson decays, the two
parameters (for each case) fit the available data quite well. For the
$D$-mesons, the values of the parameters correspond to dropping the
contribution of the color mismatched 4-quark operator altogether
\cite{BSW:87}, which is a procedure that finds some theoretical justification
in $1/N_c$ expansion arguments \cite{Buras:86}. Quite surprisingly, the recent
data on $B$-meson decays has shown that, in this case, the values of the
parameters do not obey the same rule \cite{CLEO:94}: neglecting the
contribution of the color mismatched operator cannot be used as a systematic
procedure to obtain the value of the free parameters, as it was the case for
the $D$ decays.

In view of the failure of the factorization approximation when applied to
the color mismatched operator, and the failure of the standard procedure of
dropping the contribution of that operator altogether, when applied to the
$B$-meson decays, the ``non-factorizable'' terms have to be dealt with.
Theoretical estimates, based on QCD sum rules, have been presented in
refs.~\cite{Bigi:94} and \cite{Ruckl:94}, for some decays. Here, I address
the more basic question of extracting the size of the ``non-factorizable''
terms from the data. This program was first advocated by Deshpande, Gronau
and Sutherland \cite{Desh:80}, and it has been recently applied to both
$D$- and $B$-meson decays by Cheng \cite{Cheng:94}. In this work, I
concentrate on the case of the $B$ decays; where, except for the matrix
elements of the operators with a color mismatch, the vacuum insertion
approximation can be assumed to work well \cite{Bj:90} (in particular, the
effects of inelastic final state scatterings will be neglected for the
$B$-decays considered in here). Special attention is paid to the ambiguities
in the values of the ``non-factorizable'' terms that are extracted from the
data. The results of Cheng \cite{Cheng:94} are recovered among other possible
solutions. Finally, ways of lifting these ambiguities are discussed.

\section{The ``non-factorizable'' terms in
$B \rightarrow J/\psi K$ and $B \rightarrow D \pi$}

The tree level, Cabibbo favored, hadronic weak decays of the $B$-mesons
correspond to the quark transitions $b \rightarrow c\overline{c}s$ and
$b \rightarrow c\overline{u}d$. The decay amplitudes are derived from the
effective weak Hamiltonian
\begin{eqnarray}
H_{eff} &=& \frac{G_{F}}{\sqrt{2}}  \,
[ V_{cb} V_{cs}^\ast ( C_1(\mu) {\cal O}_1^c + C_2(\mu) {\cal O}_2^c )
\nonumber\\
& & + V_{cb} V_{ud}^\ast ( C_1(\mu) {\cal O}_1^u + C_2(\mu) {\cal O}_2^u ) ] ,
\label{eq:21}
\end{eqnarray}
where
\begin{eqnarray}
{\cal O}_1^c &=&  \overline{c}_\alpha \gamma^{\mu} (1 - \gamma_5) b_\alpha \:\,
\overline{s}_\beta \gamma_{\mu} (1 - \gamma_5) c_\beta ,
\nonumber\\
{\cal O}_2^c &=& \overline{s}_\alpha \gamma^{\mu} (1 - \gamma_5)  b_\alpha \:\,
\overline{c}_\beta \gamma_{\mu}  (1 - \gamma_5) c_\beta   ,
\label{eq:22}
\end{eqnarray}
and the operators ${\cal O}_{1,2}^u$ are obtained from ${\cal O}_{1,2}^c$,
replacing $\overline{s}$ and $c$ by $\overline{d}$ and $u$, respectively.
The Wilson coefficients $C_1(\mu)$ and $C_2(\mu)$ contain the short distance
QCD corrections. In the leading logarithm approximation, they are \cite{MK:74}
$C_{1,2} = (C_+ \pm C_-)/2$, with
\begin{equation}
C_{\pm}(\mu) = \left(\frac{\alpha_s(\mu)}{\alpha_s(M_W)}\right)
^{\frac{6\gamma_{\pm}}{33-2 n_f}}
\label{eq:23}
\end{equation}
($\gamma_-=-2\gamma_+=2$; $n_f$ is the number of active flavors).
For $\Lambda_{\overline{MS}}^{(5)} = 200$ MeV \cite{Pdg:94}, and at the scale
$\mu = 5.0$ GeV, this gives
\begin{equation}
C_1 = 1.117 \hspace{.3in} C_2 = -0.266  .
\label{eq:24}
\end{equation}

For the exclusive decays $B^- \rightarrow J/\psi K^-$ or $\overline{B^0_d}
\rightarrow J/\psi \overline{K^0}$, the amplitude is
\begin{eqnarray}
\lefteqn{A(B \rightarrow J/\psi K) = }\nonumber\\
& & = - \frac{G_{F}}{\sqrt{2}} V_{cb} V_{cs}^\ast
(C_1 <J/\psi K|{\cal O}_1^c |B> + C_2 <J/\psi K|{\cal O}_2^c |B>) .
\label{eq:25}
\end{eqnarray}
The hadronic matrix element of ${\cal O}_2^c$ is calculated in the vacuum
insertion (factorization) approximation:
\begin{equation}
<J/\psi K|{\cal O}_2^c |B> =
<J/\psi| \overline{c}_\beta \gamma_{\mu} (1 - \gamma_5) c_\beta  |0>
<K| \overline{s}_\alpha \gamma^{\mu} (1 - \gamma_5) b_\alpha |B> .
\label{eq:26}
\end{equation}
Whereas, for ${\cal O}_1^c$, an additional ``non-factorizable'' term
is included that reflects the color mismatch in the $c$-$\overline{c}$
quark fields:
\begin{eqnarray}
\lefteqn{<J/\psi K|{\cal O}_1^c |B> = }\nonumber\\
& & = \frac{1}{N_c} <J/\psi| \overline{c}_\beta \gamma_{\mu}
(1 - \gamma_5) c_\beta |0> <K| \overline{s}_\alpha \gamma^{\mu}
(1 - \gamma_5) b_\alpha |B> \nonumber\\
& & + <J/\psi K|{\cal O}_1^c |B>_{non-fact.} .
\label{eq:27}
\end{eqnarray}
The factor $1/N_c$ corresponds to the projection of the two color mismatched
quark fields into a color singlet; it is included so that the
``non-factorizable'' term (that is defined by eq.~\ref{eq:27}) vanishes in the
vacuum insertion approximation. The decay amplitude can then be written as
\begin{equation}
A(B \rightarrow J/\psi K) = - \frac{G_{F}}{\sqrt{2}} V_{cb} V_{cs}^\ast M a ,
\label{eq:28}
\end{equation}
with
\begin{equation}
a = C_2 + C_1 (\frac{1}{N_c} + X) ,
\label{eq:29}
\end{equation}
\begin{equation}
M \equiv <J/\psi| \overline{c}_\beta \gamma_{\mu} (1 - \gamma_5) c_\beta  |0>
<K| \overline{s}_\alpha \gamma^{\mu} (1 - \gamma_5) b_\alpha |B>
\label{eq:210}
\end{equation}
and
\begin{equation}
X \equiv \frac{1}{M} <J/\psi K|{\cal O}_1^c |B>_{non-fact.} .
\label{eq:211}
\end{equation}
In the BSW model \cite{BSW:85}, $M=5.84 \;{\rm GeV}^3 \times
f_{\psi}/(395\; {\rm MeV})$.
With $|V_{cb}| = 0.038 \sqrt{1.63\;{\rm psec}/\tau_B}$
\cite{Ali:94}, and $f_{\psi}=395$ MeV (which cor\-res\-ponds to
$\Gamma(J/\psi \rightarrow e^+e^-)  = (5.26 \pm 0.37)$ keV \cite{Pdg:94}),
the branching ratio is
\begin{equation}
B(B \rightarrow J/\psi K) =  1.90 |a|^2 \% .
\label{eq:212}
\end{equation}
{}From the average of the experimental results \cite{Pdg:94}
\begin{eqnarray}
B(B^- \rightarrow J/\psi K^-) &=& (0.102 \pm 0.014)\% \nonumber\\
B(\overline{B^0} \rightarrow J/\psi \overline{K^0}) &=& (0.075 \pm 0.021)\% ,
\label{eq:213}
\end{eqnarray}
it follows that $|a| \simeq 0.22$, and so
\begin{equation}
X \simeq -0.29 {\rm \makebox[0.5in]{or}} 0.10
\label{eq:214}
\end{equation}
(the numerical differences with respect to the analogous results
in ref.~\cite{Cheng:94} correspond to the updated values of the
parameters that were used in here). The 2-fold ambiguity in the value of $X$,
which corresponds to the unknown sign of $a$, cannot be resolved by the
branching ratio of $B \rightarrow J/\psi K$ alone.

Proceeding in a similar way for the exclusive processes $\overline{B^0_d}
\rightarrow D^+ \pi^-$, $\overline{B^0_d} \rightarrow D^0 \pi^0$, and
$B^- \rightarrow D^0 \pi^-$, the weak decay amplitudes are
\begin{eqnarray}
A^{+-} &=& - \frac{G_{F}}{\sqrt{2}} V_{cb} V_{ud}^\ast
M_1 a_1 , \nonumber\\
A^{00} &=& - \frac{G_{F}}{\sqrt{2}} V_{cb} V_{ud}^\ast
\frac{1}{\sqrt{2}} M_2 a_2  \nonumber
\end{eqnarray}
and
\begin{equation}
A^{0-} = - \frac{G_{F}}{\sqrt{2}} V_{cb} V_{ud}^\ast
M_1 a_1 [1 + \frac{M_2}{M_1} \frac{a_2}{a_1}],
\label{eq:215}
\end{equation}
respectively (non-spectator contributions are very small and they have been
neglected). The hadronic matrix elements $M_1$ and $M_2$ are
\begin{equation}
M_1 \equiv <D^+| \overline{c}_\alpha \gamma^{\mu} (1 - \gamma_5) b_\alpha
|\overline{B^0_d}>
<\pi^-|\overline{d}_\beta \gamma_{\mu} (1 - \gamma_5) u_\beta) |0> \nonumber
\end{equation}
and
\begin{equation}
M_2 \equiv \sqrt{2} <\pi^0| \overline{d}_\alpha \gamma^{\mu} (1 - \gamma_5)
b_\alpha |\overline{B^0_d}>
<D^0| \overline{c}_\beta \gamma_{\mu}  (1 - \gamma_5) u_\beta |0> .
\label{eq:216}
\end{equation}
In the BSW model \cite{BSW:85}, $M_1=1.85\;{\rm GeV}^3$ and $M_2=2.28\;
{\rm GeV}^3$ (for $f_D = 220$ MeV). As in the previous case, the parameters
\begin{equation}
a_1 = C_1 + C_2 (\frac{1}{N_c} + X_1)  \nonumber
\end{equation}
and
\begin{equation}
a_2 = C_2 + C_1 (\frac{1}{N_c} + X_2)
\label{eq:217}
\end{equation}
include the terms
\begin{eqnarray}
X_1 &\equiv& \frac{1}{M_1} <D^+ \pi^-| {\cal O}_2^u |\overline{B^0_d}>_
{non-fact.} \nonumber\\
X_2 &\equiv& \frac{\sqrt{2}}{M_2} <D^0 \pi^0| {\cal O}_1^u |\overline{B^0_d}>_
{non-fact.} ,
\label{eq:218}
\end{eqnarray}
where the ``non-factorizable'' matrix elements are defined by
\begin{eqnarray}
<D^+ \pi^-| {\cal O}_2^u |\overline{B^0_d}> &=& \frac{1}{N_c} M_1 +
<D^+ \pi^-| {\cal O}_2^u |\overline{B^0_d}>_{non-fact.}  \nonumber\\
\sqrt{2} <D^0 \pi^0| {\cal O}_1^u |\overline{B^0_d}> &=& \frac{1}{N_c} M_2 +
\sqrt{2} <D^0 \pi^0| {\cal O}_1^u |\overline{B^0_d}>_{non-fact.}.
\label{eq:219}
\end{eqnarray}

In order to determine the values of the parameters $a_1$ and $a_2$,
the magnitudes of the decay amplitudes are extracted from the experimental
value of the corresponding branching ratios \cite{Pdg:94},
\begin{eqnarray}
B(\overline{B^0_d} \rightarrow D^+ \pi^-) &=& (0.30 \pm 0.04)\% \nonumber\\
B(\overline{B^0_d} \rightarrow D^0 \pi^0) &<&
0.048 \% \;\; (90\% C.L.) \nonumber\\
B(B^- \rightarrow D^0 \pi^-) &=& (0.53 \pm 0.05)\% ,
\label{eq:220}
\end{eqnarray}
and compared to the predictions in eq.~\ref{eq:215}. The later are the
amplitudes in the absence of final state interaction effects, and so the
comparison must be done with care (to be sure, I use the notation
${\cal A}^{+-}$, ${\cal A}^{00}$ and ${\cal A}^{0-}$ for the full amplitudes).

As for the $B \rightarrow J/\psi K$ case, it is assumed that the effects of
inelastic final state interaction scatterings are negligible. However, the
$B \rightarrow D \pi$ decays involve two isospin channels, and so an elastic
final state interaction phase, $\delta$, can appear between the two isospin
amplitudes $A_{3/2}$ and $A_{1/2}$. In general,
\begin{eqnarray}
A_{3/2} &=& |A_{3/2}| \nonumber\\
A_{1/2} &=& |A_{1/2}| e^{i\delta^\prime} ,
\label{eq:221}
\end{eqnarray}
where $\delta^\prime = \delta$ or $\delta + \pi$ (according to the relative
sign of the two amplitudes in the absence of final state interactions).
The full amplitudes ${\cal A}^{+-}$, ${\cal A}^{00}$ and ${\cal A}^{0-}$ are
related to the isospin amplitudes $A_{3/2}$ and $A_{1/2}$, in the following
way:
\begin{eqnarray}
{\cal A}^{+-} &=& \frac{1}{\sqrt{3}} A_{3/2} +
\frac{\sqrt{2}}{\sqrt{3}} A_{1/2} , \nonumber\\
{\cal A}^{00} &=& \frac{\sqrt{2}}{\sqrt{3}} A_{3/2} -
\frac{1}{\sqrt{3}} A_{1/2} , \nonumber\\
{\cal A}^{0-} &=& \sqrt{3} A_{3/2} .
\label{eq:222}
\end{eqnarray}
This allows to determine the magnitudes $|A_{1/2}|$ and $|A_{3/2}|$, as well
as $\cos \delta^\prime$, from the experimental results in eq.~\ref{eq:220}.
In particular, it follows that
\begin{equation}
\cos \delta^\prime > 0.77
\label{eq:223}
\end{equation}
The lack of a more precise value for $\delta^\prime$ is due to the fact that
only an upper-limit exists for $B(\overline{B^0_d} \rightarrow D^0 \pi^0)$.
For now, I will take $\delta^\prime = 0$ (i.e. the final state interaction
phase is either $\delta = 0$ or $\pi$). Then,
\begin{eqnarray}
|A_{3/2}| &=& \frac{1}{\sqrt{3}} |{\cal A}^{0-}| \nonumber\\
|A_{1/2}| &=& \frac{\sqrt{3}}{\sqrt{2}} ( |{\cal A}^{+-}| -
\frac{1}{3} |{\cal A}^{0-}|).
\label{eq:224}
\end{eqnarray}
Since the decay amplitudes in the absence of final state interactions are
\begin{eqnarray}
A^{+-} &=& \frac{1}{\sqrt{3}} |A_{3/2}| \pm
\frac{\sqrt{2}}{\sqrt{3}} |A_{1/2}| , \nonumber\\
A^{00} &=& \frac{\sqrt{2}}{\sqrt{3}} |A_{3/2}| \mp
\frac{1}{\sqrt{3}} |A_{1/2}| , \nonumber\\
A^{0-} &=& \sqrt{3} |A_{3/2}|
\label{eq:225}
\end{eqnarray}
(the two signs correspond to $\delta^\prime = \delta$ or $\delta + \pi$),
it follows that
\begin{equation}
|A^{+-}| = |{\cal A}^{+-}| {\rm \makebox[0.4in]{}} \frac{A^{0-}}{A^{+-}} =
\frac{|{\cal A}^{0-}|}{|{\cal A}^{+-}|} ,
\label{eq:226}
\end{equation}
for $\delta = 0$; and
\begin{equation}
|A^{+-}| = |\frac{2}{3} |{\cal A}^{0-}| - |{\cal A}^{+-}||
{\rm \makebox[0.4in]{}}
\frac{A^{0-}}{A^{+-}} = \frac{|{\cal A}^{0-}|}{\frac{2}{3}|{\cal A}^{0-}| -
|{\cal A}^{+-}|} ,
\label{eq:227}
\end{equation}
for $\delta = \pi$.
The predictions of eq.~\ref{eq:215}, in terms of the parameters $a_1$ and
$a_2$, are replaced on the LHS of these equations; whereas the experimental
input from the branching ratios is used for the RHS. If the experimental
result of eq.~\ref{eq:223} is interpreted as showing a negligible phase shift
from the final state interaction effects, eq.~\ref{eq:226} gives
(with $|V_{cb}|$ as before)
\begin{equation}
|a_1| \simeq 1.07 \hspace{.2in} 1 + 1.23 \frac{a_2}{a_1} \simeq 1.33 ;
\label{eq:228}
\end{equation}
and the size of the ``non-factorizable'' terms is then
\begin{equation}
X_1 \simeq -0.17 \hspace{.2in} X_2 \simeq 0.16
\label{eq:229a}
\end{equation}
or
\begin{equation}
X_1 \simeq 7.90  \hspace{.2in} X_2 \simeq -0.35
\label{eq:229b}
\end{equation}
(the positive value for $X_2$ corresponds to the result in
ref.~\cite{Cheng:94}).
Alternatively, the data can be interpreted as showing a maximal phase shift.
Then eq.~\ref{eq:227} gives
\begin{equation}
|a_1| \simeq 0.12 \hspace{.2in} 1 + 1.23 \frac{a_2}{a_1} \simeq -11.63 ;
\label{eq:230}
\end{equation}
and so
\begin{equation}
X_1 \simeq 1.70 \hspace{.2in} X_2 \simeq -0.61
\label{eq:231a}
\end{equation}
or
\begin{equation}
X_1 \simeq 2.16  \hspace{.2in} X_2 \simeq 0.52 .
\label{eq:231b}
\end{equation}
The 4-fold ambiguity corresponds to the fact that the final state interaction
phase can only be determined {\em modulo} $\pi$, and the sign of $a_1$ cannot
be determined from the branching ratios in eq.~\ref{eq:220}.

At this point, a word should be said about the uncertainties in the results
that were presented. The derivation of the parameters $|a|$, $|a_1|$ and
$|a_2|$
suffers from the experimental errors in the branching ratios (in particular,
$B(\overline{B^0_d} \rightarrow D^0 \pi^0)$ is still missing), and in
$|V_{cb}|$. These will improve with more accumulated data; which will also
allow to derive the hadronic matrix elements of the bilinear quark operators
from the semileptonic branching ratios (see the tests of factorization in
ref.~\cite{CLEO:94}, for example). At present, the use of the BSW model
{}~\cite{BSW:85} entails an uncertainty that is hard to quantify.
The derivation of the terms $X$, $X_1$ and $X_2$ suffers from the additional
uncertainty on the Wilson coefficients: in particular, the value chosen for
the scale $\mu$ is important  \cite{NLO:94}. The positive solution for $X$ and
the negative solution for $X_1$ are the most sensitive as they involve
somewhat delicate cancellations; for $\mu$ in the range $4.5$ to $5.5$ GeV,
they oscillate by as much as $25\%$ (whereas the other results vary by not
more than $10\%$). I have taken $\mu = 5.0$ GeV, which is approximately the
constituent $b$-quark mass.

It must be stressed that the vacuum insertion result has been assumed to
work well for the operators with the correct color assignments (see for
example eq.~\ref{eq:26}), and the inelastic final state scatterings have been
deemed negligible. This is necessary in order to be able to derive the
size of the ``non-factorizable'' terms from the data. In ref.~\cite{Bj:90}, it
is argued, on the basis of color transparency, that these assumptions are
expected to hold for the low-multiplicity $B$ decays. The argument is less
reliable for the case of the $D$ decays, and so they have not been considered
in here.

\section{Resolving the ambiguities}

\subsection{``Non-factorizable'' terms in $B \rightarrow J/\psi K$}

The sign of the parameter $a$, that appears in the $B \rightarrow J/\psi K$
amplitude of eq.~\ref{eq:28}, can be determined from the interference between
the short distance contribution to $B \rightarrow K l^+ l^-$, and the long
distance contribution due to $B \rightarrow K J/\psi \rightarrow K l^+ l^-$.
The short distance amplitude is derived from the effective weak Hamiltonian
in eq.~\ref{eq:21} (the operators ${\cal O}_{1,2}^c$ contribute
at the 1-loop level), with the additional electroweak terms:
\begin{equation}
H_{eff}^\prime = \frac{G_{F}}{\sqrt{2}} V_{tb} V_{ts}^\ast
\sum_{i=7,8,9} C_i(\mu) {\cal O}_i ,
\label{eq:31}
\end{equation}
where the operators
\begin{eqnarray}
{\cal O}_7 &=&  \frac{e}{8\pi^2} m_b \overline{s}_\alpha \sigma^{\mu\nu}
(1 + \gamma_5) b_\alpha F_{\mu\nu}
\nonumber\\
{\cal O}_8 &=&  \frac{\alpha}{2\pi} \overline{s}_\alpha \gamma^\mu
(1 - \gamma_5) b_\alpha \overline{l} \gamma_\mu l
\nonumber\\
{\cal O}_9 &=&  \frac{\alpha}{2\pi} \overline{s}_\alpha \gamma^\mu
(1 - \gamma_5) b_\alpha \overline{l} \gamma_\mu \gamma_5 l
\label{eq:32}
\end{eqnarray}
contribute to $B \rightarrow K l^+ l^-$ at tree level. For $m_t = 175$ GeV,
the Wilson coefficients in eq.~\ref{eq:31}, in the leading logarithm
approximation, are \cite{Grin:89} $C_7 = 0.326$, $C_8 = -3.752$ and
$C_9 = 4.581$. The decay amplitude is
\begin{eqnarray}
A &=& \frac{G_{F}\alpha}{\sqrt{2}\pi} V_{tb} V_{ts}^\ast
[- <K| \overline{s} \gamma^\mu b |B> \frac{1}{2}
( C_{8eff} \overline{u}_l \gamma_\mu v_{\overline{l}}
+ C_9 \overline{u}_l \gamma_\mu \gamma_5 v_{\overline{l}}) \nonumber\\
& & + <K|\overline{s} i \sigma^{\mu\nu} q_{\nu} (1+\gamma_5) b |B>
C_7  m_b \frac{1}{q^2} \overline{u}_l \gamma_\mu v_{\overline{l}}]
\label{eq:33}
\end{eqnarray}
($q \equiv p_B - p_K$). The factor
\begin{equation}
C_{8eff} = C_8 - (3C_2 + C_1) g(\frac{4m_c^2}{q^2},\frac{m_c^2}{m_b^2}) +
3 a g_{LD}
\label{eq:34}
\end{equation}
includes the contribution
\begin{eqnarray}
g(x,y) &=& - \frac{4}{9} \ln y + \frac{4}{9} x + \frac{8}{27}
- \frac{8}{9} (1+\frac{1}{2}x) \sqrt{x-1} \arctan \frac{1}{\sqrt{x-1}}
\theta(x-1) \nonumber\\
& & - \frac{4}{9} (1+\frac{1}{2}x) \sqrt{1-x} (\ln \frac{1+\sqrt{1-x}}
{1-\sqrt{1-x}} + i \pi) \theta(1-x)
\label{eq:35}
\end{eqnarray}
from the operators ${\cal O}_{1,2}^c$, and the long distance contribution
\cite{Ali:91}
\begin{equation}
g_{LD} = \frac{3\pi}{\alpha^2} \sum_{V=J/\psi,\psi^\prime}
\frac{m_V \Gamma (V \rightarrow l^+ l^-)}{q^2 - m_V^2 + i m_V \Gamma_V}
\label{eq:36}
\end{equation}
from the $J/\psi$ and $\psi^\prime$ resonances. The parameter $a$ that
multiplies $g_{LD}$ is the same as in eq.~\ref{eq:28} (for simplicity, I have
taken the same parameter for both the $J/\psi$ and the $\psi^\prime$
resonances,
but this is not necessary); the relative sign between the long distance and
short distance contributions is well determined \cite{Lim:89}, up to the sign
of $a$. The hadronic matrix elements are parameterized by
\begin{eqnarray}
<K| \overline{s} \gamma^\mu b |B> &=& (p_B + p_K)^\mu f_+(q^2)
+ q^\mu f_-(q^2) ,
\nonumber\\
<K|\overline{s} i \sigma^{\mu\nu} (1+\gamma_5) b |B> &=&
s(q^2) [(p_B + p_K)^\mu q^\nu - (p_B + p_K)^\nu q^\mu \nonumber\\
& & + i\epsilon_{\mu\nu\alpha\beta} (p_B + p_K)^\alpha q^\beta] ,
\label{eq:37}
\end{eqnarray}
where, in the static $b$-quark limit \cite{Isg:90}, $s = -(f_+ - f_-)/2m_B$.
The modified BSW model \cite{BSW:85} gives
\begin{eqnarray}
f_+(q^2) &=& \frac{h_0}{1-\frac{q^2}{m_{0^+}^2}}
 \frac{1}{1-\frac{q^2}{(m_B + m_K)^2}} , \nonumber\\
f_-(q^2) &=& - f_+(q^2) \frac{m_B - m_K}{m_B + m_K} ,
\label{eq:37b}
\end{eqnarray}
with $h_0 = 0.379$ and $m_{0^+} = 5.89$ GeV.
The differential branching ratio is then
\begin{eqnarray}
\frac{1}{\Gamma}\frac{d\Gamma}{dz} &=&  \frac{1}{48} \tau_B G_F^2 \alpha^2
|V_{tb}V_{ts}^\ast|^2 (\frac{m_B}{2\pi})^5 (1-z)^3 f_+^2
\nonumber\\
& & \times (|C_9|^2 + |C_{8eff} + 2 C_7|^2)
\label{eq:38}
\end{eqnarray}
($z\equiv q^2/m_B^2$), where the lepton and kaon masses were neglected.
This is shown in fig. 1 for $a$ positive and negative. Studying the region
of the interference between the short and the long distance contributions
will allow to determine the sign of $a$, and resolve the ambiguity in
eq.~\ref{eq:214}. At present, the necessary sensitivity has not been reached
yet, and only an upper limit exists on the non-resonant $B \rightarrow K l^+
l^-$ decays \cite{CDF:94}.

\subsection{``Non-factorizable'' terms in $B \rightarrow D \pi$}

The fortunate interference that allows to determine the sign of $a$ is quite
unique, and no similar effect appears for the decays of the type $b
\rightarrow c\overline{u}d$, that would allow to determine the sign of $a_1$
in eq.~\ref{eq:215}. As for the final state interaction phase $\delta$ in
eq.~\ref{eq:221}, it is known \cite{Kam:86} that it should be the same phase
that appears in $D$-$\pi$ elastic scattering, at the energy $E_{c.m.}=m_B$.
But this is of little use in determining its value. Indeed, it is hard to
think of an experimental test that would lift the 4-fold ambiguity in the
values of the two ``non-factorizable'' terms in the $B \rightarrow D \pi$
decays. On the other hand, it has been assumed throughout the analysis that
the vacuum insertion result is a good approximation for the matrix elements
of the operators with the correct color assignments. If it is further assumed
that the vacuum insertion result should provide a first order approximation for
the matrix elements of the color mismatched operators, then the $X$-terms
should not be larger than unity. In particular, the solution for the
$B \rightarrow D \pi$ decays is $X_1 \simeq -0.17$ and $X_2 \simeq 0.16$,
as in eq.~\ref{eq:229a}.

Although there is no reason to expect that this is so (the arguments in
ref.~\cite{Bj:90}, for example, cannot be extended to the case of the
``non-facto\-ri\-za\-ble'' terms), it should be pointed out that the solutions
with small $X$-terms tend to agree with the values predicted by the theoretical
calculations that are presently available. Using QCD sum rules techniques, it
has been predicted that $X_1 \simeq -0.33$ \cite{Bigi:94} (and, for the case
of $B \rightarrow J/\psi K$, $X$ is between $-0.30$ and $-0.15$
\cite{Ruckl:94}). Similar results have been obtained for the
``non-factorizable'' term in the amplitude for $B^0-\overline{B^0}$ mixing. The
hadronic matrix element in the mixing amplitude is
\begin{equation}
M_{mix.} \equiv <B^0_q| \overline{q}_\alpha \gamma^\mu (1 - \gamma_5) b_\alpha
\overline{q}_\beta \gamma^\mu (1 - \gamma_5) b_\beta |\overline{B^0_q}> .
\label{eq:39}
\end{equation}
Proceeding in a similar way as for the decay amplitudes, a
``non-facto\-ri\-za\-ble'' term is defined by
\begin{eqnarray}
M_{mix.} &=& 2 (1 + \frac{1}{N_c}) m_B^2 f_B^2  \nonumber\\
& & + <B^0_q| \overline{q}_\alpha \gamma^\mu (1 - \gamma_5) b_\alpha
\overline{q}_\beta \gamma^\mu (1 - \gamma_5) b_\beta
|\overline{B^0_q}>_{non-fact.}
\label{eq:310}
\end{eqnarray}
(where $<B^0_q(p)|\overline{q} \gamma^\mu (1 - \gamma_5) b|0> = -i f_B p^\mu$).
The deviation from the vacuum insertion result (the first terms on the RHS of
eq.~\ref{eq:310}) is parameterized by the bag parameter $B_B$. Here,
\begin{equation}
B_B = 1 + \frac{1}{(1 + \frac{1}{N_c})} X_{mix.} ,
\label{eq:311}
\end{equation}
where
\begin{equation}
X_{mix.} \equiv \frac{<B^0_q| \overline{q}_\alpha \gamma^\mu (1 - \gamma_5)
b_\alpha \overline{q}_\beta \gamma^\mu (1 - \gamma_5) b_\beta
|\overline{B^0_q}>_{non-fact.}}{2 m_B^2 f_B^2} .
\label{eq:312}
\end{equation}
At present, the magnitude of $B_B$ cannot be derived from the data on
$B^0_d-\overline{B^0_d}$ mixing, because one lacks a precise determination of
$f_B$ and of $|V_{td}|$ (for $B^0_s-\overline{B^0_s}$ mixing, the CKM parameter
is better known, but the strength of the mixing has not been determined
experimentally). Using the lattice ($B_B = 1.2 \pm 0.2$) \cite{Shi:94} and
the QCD sum rules ($B_B = 1.0 \pm 0.15$) \cite{Nar:94} estimates for the
matrix element in the mixing, it follows that
\begin{equation}
0 \leq X_{mix.} \leq 0.53 {\rm \makebox[0.6in]{and}}
-0.2 \leq X_{mix.} \leq 0.2 ,
\label{eq:313}
\end{equation}
respectively.

\section{Conclusion}

The size of the ``non-factorizable'' terms in the amplitude for the decays
$B \rightarrow J/\psi K$ and $B \rightarrow D \pi$ was derived from the
experimental value of the corresponding branching ratios. The results can only
be determined up to a discrete ambiguity. In the case of $B \rightarrow J/\psi
K$, the 2-fold ambiguity can be lifted by determining the sign of the
interference between the short distance contribution to $B \rightarrow
K l^+ l^-$, and the long distance contribution due to $B \rightarrow K J/\psi
\rightarrow K l^+ l^-$. A similar ambiguity appears in the case of the
$B \rightarrow D \pi$ decays; because the final state interaction
elastic phase between the two isospin amplitudes can only be determined
{\em modulo} $\pi$, the ambiguity becomes 4-fold. Contrary to the previous
case, there is no simple way to determine the correct solution experimentally.
However, all but one of the solutions indicates large ``non-factorizable''
terms that would indicate a breakdown of the vacuum insertion approximation
when applied to the color mismatched operators.
The analysis that was presented can be improved with future experimental
results, in particular, with a measurement of the branching ratio for
$\overline{B^0_d} \rightarrow D^0 \pi^0$. Also, other $B$-decays, similar to
the ones shown in here, can be considered; the ambiguities in the values of the
``non-factorizable'' terms that are extracted from the data will appear in the
same fashion.

\section*{}

I would like to thank A. Buras for a very useful discussion, and H.-Y. Cheng
for carefully reading the manuscript and for his comments and corrections.
This work was partly supported by the Natural Science and Engineering Research
Council of Canada.

\pagebreak
\pagestyle{empty}
\section*{Figure Caption}
\begin{tabbing}
\=Figure 1: \= Differential branching ratio for $B \rightarrow K l^+ l^-$
($z \equiv (p_B - p_K)^2/m_B^2$).\\
\>	\> The full line corresponds to the long distance
contribution alone,\\
\>	\> whereas the other curves include the short distance
contribution:\\
\>	\>with $a > 0$ (dashed line) and $a < 0$ (dotted line).
\end{tabbing}

\begin{thebibliography}{Abcd:efgh}
\bibitem{NLO:94}
The strong $\mu$ dependence persists, when QCD corrections are included
at the next-to-leading order \cite{Per:94} \cite{Buras:94}. In ref.
\cite{Per:94}, it is shown that the higher order corrections must be
included as part of a simultaneous expansion in $C_2 + C_1/N_c$ and
$\alpha_s$, to avoid unphysical results; no large deviation from the
LLA result is apparent in the $B \rightarrow J/\psi + X_s$ decay. A
different approach is advocated in ref. \cite{Buras:94} for the exclusive
decays. Using the NLO results in there would lead to a large renormalization
scheme dependence in the ``non-factorizable'' terms.
\end{thebibliography}
\end{document}